\documentstyle[aps,epsfig]{revtex}

\begin{document}

\title{ Nucleonic resonance excitations 
with linearly polarized photon in $\gamma p\to \omega p$ }

\author{Qiang Zhao\thanks{Present address: Dept. of Physics, Univ. of Surrey, 
Guildford, GU2 7XH, UK; 
Email: qiang.zhao@surrey.ac.uk.} }  
\address{ Institut de Physique Nucl\'eaire, F-91406 Orsay 
Cedex, France } 

\date{\today}

\maketitle  
  
\begin{abstract}
In this work, an improved quark model approach 
to the $\omega$ meson photo-production 
with an effective Lagrangian is presented. The {\it t}-channel 
{\it natural}-parity 
 exchange is taken into account through the Pomeron exchange,  
while the {\it unnatural}-parity exchange is described by the $\pi^0$ 
exchange. 
With a very limited number of parameters, 
the available experimental data
in the low energy regime can be consistently accounted for. 
We find that 
the beam polarization observables show sensitivities to some 
 {\it s}-channel individual resonances in the $SU(6)\otimes O(3)$
quark model symmetry limit. 
Especially, the two resonances $P_{13}(1720)$ and 
$F_{15}(1680)$, which belong to the representation $[{\bf 56, ^2 8}, 2, 2, J]$, 
have dominant contributions over other excited states. 
Concerning the essential motivation of searching for ``missing resonances" 
in meson photo-production, this approach provides a feasible framework, 
on which systematic investigations can be done.

\end{abstract}

\vskip 1.cm

{PACS number(s): 13.60.-r, 13.60.Le, 14.20.Gk, 12.39.-x }

\newpage

\section{ Introduction}

The non-relativistic constituent quark model (NRCQM)~\cite{NRCQM-1,NRCQM-2} 
predicts a much richer baryon spectroscopy 
than observed in $\pi N\to \pi N$ scatterings. 
This arises the efforts 
of searching for those so-called ``missing resonances" in other 
meson coupled channels, for example, 
$\eta N$, $\Delta \pi$, $K\Lambda$, $K \Sigma$, 
$\rho N$, $\omega N$, $\phi N$, etc., 
to which those resonances might have stronger couplings 
such that can be detected in experiment.

The availabilities of high-intensity electron and photon facilities 
from JLAB, ELSA, ERSF, SPring-8
provide accesses to exciting the nucleons with the clean
electromagnetic probes. Abundant informations of 
the internal structures of nucleons and their
excited states can be then derived in their consequent meson decays. 
Namely, the meson photo-production 
{\it via} resonance excitations might be able to provide an ideal tool 
for the essential motivations~\cite{jlab-91-024,jlab-94-109,jlab-98-109,jlab-99-013}.
It can be anticipated that in the near future, a large database for the 
study of baryon structures as well as their couplings to 
various meson channels can be established. 
Theoretical interests are also revived in this QCD subfield,
which is non-perturbative in nature. 
Empirical or phenomenological prescriptions with QCD-inspired ingredients
are therefore developed 
in history~\cite{close-li,li-close,capstick92,capstick94,capstick98}, 
which will be tested by the forthcoming data.

In this work, we will concentrate on the exclusive
$\omega$ meson photo-production, $\gamma p\to \omega p$. 
Exploring over the available, but very limited, data 
in literatures~\cite{klein,ABBHHM68,ballam73,old-data}, 
we aim at establishing a consistent framework for the study of the
intermediate resonance excitations in this channel. 
Since large degrees of freedom will appear at the resonance region
~\footnote{In principle, for each resonance 24 numbers (12 independent amplitudes)
are required to determine its 
transition matrix elements in vector meson photo-production.
In pseudoscalar meson photo-production, 8 numbers are required. },
we adopt the quark model description of the
baryon states for the nucleons and resonances
in this reaction, which will significantly reduce
the number of free parameters.

In Ref.~\cite{plb98,prc98}, a quark model for the resonance excitations
has been developed. 
Taking the advantages of the self-consistent quark model framework, 
the authors introduced an effective Lagrangian for the 
quark-vector-meson vertex, where the vector meson was treated as a point-like
elementary particle. 
At quark level, only two universal parameters for the 
vector and tensor couplings would appear in the resonance excitation terms.

However, that approach needs to be improved 
since the {\it t}-channel {\it natural}-parity
exchange has not been taken into account there. 
Only the {\it unnatural}-parity $\pi^0$ exchange was included
to describe the forward peaking phenomenon in $\gamma p\to \omega p$. 
Due to the absence of the {\it natural}-parity exchange,
the interferences between the {\it unnatural}-parity exchange and  
the {\it s}- and {\it u}-channels 
might be over-estimated in some polarization observables. 

In this improved version, the {\it natural}-parity exchange will be included 
through a Pomeron exchange phenomenology. 
In our preliminary study~\cite{hadron99}, we have shown 
that the introduction of the {\it natural}-parity exchange 
can improve the description of the cross sections very significantly. 
Here, for the first time, a quantitative results is derived following a
parameter constraint scheme, although more rigorous constraints to the
parameters are expected from the forthcoming data from GRAAL and JLAB. 
The roles played by the {\it t}-channel {\it natural} and {\it unnatural}-parity
exchanges are emphasized since a reliable estimation of this part
should be the prerequisite for further study relevant to the nucleonic resonance
excitations.
One will see as follows, over a large energy regions, the interferences 
between the {\it natural} and {\it unnatural}-parity exchanges 
provide good constraints to the {\it t}-channel processes, 
which makes it possible to highlight the roles played by the intermediate resonances.
We also show that
the {\it natural}-parity exchange
is neccessary to account for the missing cross sections 
when $E_\gamma>\sim 2.2$ GeV.

In section II, an introduction of the quark-vector-meson coupling 
is presented in the 
$SU(6)\otimes O(3)$ quark model symmetry limit. 
Also, the {\it t}-channel Pomeron exchange model is summarized.
In section III, the numerical study of the the resonance effects
 in the cross sections and beam polarization observables are presented. 
The conclusions and discussions are given in section IV.

\section{ $\omega$ meson photo-production amplitudes }

In this section, the three contributing processes in the $\omega$ meson 
photo-production are summarized.

\subsection{ {\it s}- and {\it u}-channel transition amplitudes }
 
In the $SU(6)\otimes O(3)$ symmetry limit, 
the quark-vector-meson coupling is described 
by the effective Lagrangian~\cite{plb98,prc98}: 
\begin{equation} \label{Lagrangian}  
L_{eff}=-\overline{\psi}\gamma_\mu p^\mu\psi+\overline{\psi}\gamma_
\mu e_qA^\mu\psi +\overline{\psi}(a\gamma_\mu +  
\frac{ib\sigma_{\mu\nu}q^\nu}{2m_q}) \phi^\mu_m \psi,
\end{equation}  
where $\psi$ and $\overline{\psi}$ represent the quark and antiquark
field, respectively, and $\phi^\mu_m $ denotes the vector meson 
field. The two parameters, 
$a$ and $b$ represent the vector and tensor couplings 
of the quark to the vector meson, respectively. 
$m_q$ and $e_q$ are the mass and charge of the constituent quark, 
respectively. For the quark model parameters, $m_q$ and $\alpha$ 
(harmonic oscillator potential strength), the commonly used values 
330 and 385 MeV are adopted, respectively. It should be noted 
that the values for $\alpha$ are different from that used in the previous 
study~\cite{prc98,hadron99}. 
In this work, an overall investigation favors smaller values.

With such an effective coupling, the transition amplitudes for the {\it s}- 
and {\it u}-channel can be explicitly derived~\cite{prc98}. 
The intermediate resonances will be 
introduced in the harmonic oscillator basis. 
Resonances with masses less than 2 GeV are assigned to the
states of $n\le 2$ in the harmonic oscillator basis, therefore, can be
taken into account explicitly in the NRCQM.  
Then, in the helicity space 
the amplitude for each resonance with 
spin $J$ can be expressed as: 
\begin{equation}
H^J_{a\lambda_v}=\frac{2 M_R}{s-M_R^2+iM_R\Gamma (q)}
\sum_{\Lambda_f}d^J_{\Lambda_f,\Lambda_i}(\theta)  
A^V_{\Lambda_f}A^\gamma_{\Lambda_i},    
\label{ampli-J}
\end{equation}    
where $s$ is the c.m. energy square and $M_R$ is the resonance mass. 
$\Gamma ({\bf q})$ is the momentum dependence of the resonance width, 
which has been given in Ref.~\cite{prc98}.
$\Lambda_f$ and $\Lambda_i$ 
are defined as,  
$\Lambda_f\equiv\lambda_v-\lambda_2$, and $\Lambda_i\equiv\lambda-\lambda_1$. 
Here $\lambda_\gamma=\pm 1$, $\lambda_v=0, \pm 1$, 
$\lambda_1=\pm 1/2$ and $\lambda_2=\pm 1/2$ denote 
the helicities of the photon, $\omega$ meson, initial
and final state nucleon, respectively.  
$A^\gamma_{\Lambda_i}$ and $A^V_{\Lambda_f}$ are the helicity 
amplitudes for the photon excitation and vector meson decay of  
an intermediate resonance, respectively. 
 $d^J_{\Lambda_f,\Lambda_i}(\theta)$ is the Wigner-$d$ function 
which rotate the vector $A^\gamma_{\Lambda_i}$ from the initial 
state coordinate system to the final state coordinate system.
States with $n> 2$ are treated as degenerate.

In the case of the $\omega$ meson photo-production, the {\it t}-channel 
and the seagull term from Eq.~\ref{Lagrangian} vanish since they are proportional to the charge 
of the outgoing vector meson, therefore vanish in the neutral vector 
meson productions. 
The nucleon pole terms have been explicitly included~\cite{prc98}.

\subsection{ {\it t}-channel {\it natural} and {\it unnatural}
parity exchange amplitudes}

We use the Pomeron exchange model by Donnachie and 
Landshoff~\cite{Donnachie,Laget,Haakman,lee} to account for the
{\it natural}-parity exchange. 
In such a model, the Pomeron mediates the long range interaction 
between a confined quark and a nucleon, 
and behaves rather like a  $C=+1$ isoscalar photon.
In Ref.~\cite{phi99}, the formula for the $\phi$ meson photo-production 
off proton have been derived. Based on the Regge phenomenology and 
the $SU(3)$ flavor symmetry, 
the same picture can be applied 
to the $\omega$ and $\rho$ meson photo-production. 
Here, we summarize the main points as follows:

The Pomeron-nucleon coupling is described by the 
vertex: 
\begin{eqnarray}
F_{\mu}(t)= 3\beta_0\gamma_{\mu}F_1(t),
\end{eqnarray}
where $\beta_0$ is the coupling strength of the single 
Pomeron to a light constituent quark ($u$, $d$, $s$), 
and is the only adjustable parameter in this Pomeron exchange model.
$F_1(t)$  is the isoscalar
nucleon electromagnetic form factor, and has the following 
expression: 
\begin{eqnarray}
F_1(t)= \frac{(4M^2_N-2.8t)}{(4M^2_N-t)(1-t/0.7)^2}.
\end{eqnarray} 
For the $\gamma \omega {\mathcal P}$ vertex 
(${\mathcal P}$ denotes the Pomeron.), 
the lowest order diagram
for the quark pair creation in Ref.~\cite{lee} is used for the 
$q\overline{q}$ creation, but has been extrapolated to the 
limit of $Q^2=0$, namely, the process with real photons. 
Meanwhile, a bare photon vertex is introduced for the quark-photon interaction, 
which has the same form as the quark-photon coupling in Eq.~\ref{Lagrangian}.

The ``on-shell approximation" is adopted for the 
$q\overline{q}\omega$ vertex $V_\nu$: 
\begin{eqnarray}
V_\nu(p-\frac 12 q, p+\frac 12 q)=f_\omega M_\omega\gamma_\nu \ ,
\end{eqnarray}
where $f_\omega$ represents the coupling strength and is fixed 
by the $\omega$ radiative decay width 
$\Gamma_{\omega\to e^+e^-}=0.59$ keV~\cite{PDG98}
with the following relation:
\begin{eqnarray}
\label{phidecay}
\Gamma_{\omega \to e^+e^-}=\frac{8\pi \alpha^2_e \hat{e}_q^2}{3}
(\frac{f^2_\omega}{M_\omega}) \ ,
\end{eqnarray}
where $\hat{e}_q$ is the charge factor of quark $q$
in terms of the charge of electron.

Therefore, the current matrix element can be written as:
\begin{eqnarray}
\label{pomeron}
\langle p_fm_f, q\lambda_v|J^\mu|p_im_i \rangle
=2\beta_0 t^{\mu\alpha\nu}(k,q)\varepsilon_{m \nu}(q){\mathcal G_P}(s,t)
\overline{u}(p_f)F_\alpha(t)u(p_i) \ ,
\end{eqnarray}
where $u(p_i)$ and $\overline{u}(p_f)$ are the initial and final 
state Dirac spinors of the protons with four-momenta $p_i$
and $p_f$, respectively.  
$k$ and $q$ are the four-momenta of the incoming photon and 
outgoing $\omega$ meson, respectively. 
$\varepsilon_{m \nu}$ is the polarization vector of the produced 
$\omega$ meson, and the factor 2 counts the equivalent contributions
from Pomeron-quark and Pomeron-anti-quark interactions.
${\mathcal G_P}(s,t)$ is related to the Regge trajectory of the 
Pomeron 
and has the form:
\begin{eqnarray}
{\mathcal G_P}(s,t)=-i(\alpha^\prime s)^{\alpha(t)-1} \ ,
\end{eqnarray}
where $\alpha(t)=1+\epsilon+\alpha^\prime t$ is the Regge trajectory
of the Pomeron. The form factor 
$\mu^2_0/(\mu^2_0+p^2)$ is introduced for each off-shell quark line with 
four-momentum $p$. The same parameter values 
as used in Ref.~\cite{lee} are adopted: 
$\epsilon=0.08$, $\alpha^\prime=0.25$ GeV$^{-2}$, $\mu_0=1.2$ GeV. 

In Eq.~\ref{pomeron}, $t^{\mu\alpha\nu}(k,q)$ represents 
the loop tensor and it has the following
expression for the contributing terms:
\begin{eqnarray}
\label{gaug-1}
t^{\mu\alpha\nu}(k,q)=(k+q)^\alpha g^{\mu\nu}-2k^\nu g^{\alpha\mu} \ .
\end{eqnarray}
To preserve gauge invariance, we have adopted
the transformation given in Ref.~\cite{Titov}. 

In Ref.~\cite{phi99}, we have shown that the polarization 
observable asymmetries at large angles depends more on the spin structures
of the {\it s}- and {\it u}-channel transition amplitudes rather than 
on the Pomeron exchange amplitude structure even the Pomeron exchange 
plays dominant role there. This feature suggests that the Pomeron phenomenology
does not influence the large angle behaviors significantly. 
Otherwise, the large angle behaviors might be ``distorted"
in the polarization observables.

The formulae for the $\pi^0$ exchange terms in the $\omega$ meson 
photo-production has been derived in Ref.~\cite{prc98}. Since the 
$\pi NN$ and $\omega\pi\gamma$ couplings are quite well known, 
the only parameters left is the form factor from the quark model,
which is to describe the 
combined effects from the two non-perturbative vertices. 
Taking into account the diffractive Pomeron exchange contribution, 
we have to adopt a smaller value $\alpha_\pi=290$ MeV for the form 
factor $e^{-({\bf q}-{\bf k})^2/6\alpha^2_\pi}$. The determination 
of $\alpha_\pi$ will be discussed later.

\section{ Numerical results }

In this section, the unpolarized differential cross sections,
density matrix elements of the vector meson 
and beam polarization observables 
in the $\omega$ meson 
photo-production are investigated. 
In $\gamma p\to \omega p$, the isospin 
conservation eliminates contributions from intermediate states 
with isospin 3/2. 
In addition, the Moorhouse selection rules~\cite{moorhouse}
 eliminate those nucleonic 
states of representation $[{\bf 70, ^4 8}, n, L, J]$ in the NRCQM 
from contributing.
Therefore, only 8 intermediate nucleonic resonances contribute
for $n\le 2$ in the harmonic oscillator basis, 
i.e. $P_{11}(1440)$, 
$S_{11}(1535)$, $D_{13}(1520)$, $P_{13}(1720)$, $F_{15}(1680)$, $P_{11}(1710)$,
$P_{13}(1900)$, $F_{15}(2000)$~\cite{PDG98}. 
In the notation for the quark model representation, 
the first three bold digitals denote the dimensions of the $SU(6)$ (spin-flavor), 
$SU(2)$ (spin), and $SU(3)$ (flavor) groups, respectively. 
$n$, $L$, and $J$ denote the main quantum number, total orbital angular momentum, 
and total spin of a three-quark baryon system in the harmonic oscillator basis.
In this convention, one can express, for example, the proton and neutron 
groundstate as $[{\bf 56, ^2 8}, 0, 0, 1/2]$.

Notice that Y. Oh {\it et al} recently studied the resonance excitations
in the $\omega$ meson photo-production~\cite{oh}. 
In their approach, the $N^*$ resonances are introduced by employing the 
quark model predictions~\cite{capstick92,capstick94} 
on the resonance photo-excitations and 
their decays into $\omega N$. Therefore, only those resonances with 
masses above the $\omega N$ threshold can be taken into account.
In our model, the low-lying resonances, for example, $S_{11}(1535)$ and $D_{13}(1520)$, 
are found still have considerable contributions, which can be seen 
in the beam polarization asymmetries. We expect that forthcoming data should 
be helpful to clarify the roles played by those below-threshold resonances.

Before proceeding to the details of the numerical calculations, 
we summarize the scheme of determining the parameters as follows.

Since it has been well-established that the high-energy $\omega$ 
production is dominated by the diffractive {\it natural}-parity exchange, 
we take the advantages of high energy measurements for the $\omega$ 
meson photoproduction by applying this Pomeron exchange 
model to $\sim$ 10 GeV, where other processes become negligible.
Then, the parameter $\beta_0=1.27$ can be derived.
In fact, this parameter is the same as used in Ref.~\cite{phi99}, 
which shows that the $SU(3)$ flavor symmetry has been conserved for the 
light quark sector.
  
Above the resonance region ($E_\gamma \ge 2.2$ GeV), the interferences
between the {\it t}-channel {\it natural} and {\it unnatural}-parity 
exchanges can be seen at forward angles up to $\sim$ 5 GeV~\cite{ballam73}, 
where the {\it s}- and 
{\it u}-channel contributions are negligible. 
It provides 
an important constraints to the $\pi^0$ exchange terms. 
Therefore, the parameter $\alpha_\pi=290$ 
can be determined.

Concerning the two parameters introduced for the {\it s}- and {\it u}-channels,
we require that a flattened behavior appearing at large angles for the 
differential cross section as shown by the old data~\cite{ABBHHM68}, as
well as the preliminary data from 
SAPHIR~\cite{klein} and CLAS~\cite{clas99}. 
Meanwhile, the total cross 
sections~\cite{ABBHHM68,ballam73,old-data} can provide constraints to the magnitudes
of the parameters. 
Here, we find that with $a=-2.5$, and $b=-6.5$, 
a flattened structure is produced 
at large angles, and the energy-evolutions of the cross sections 
can be reasonably reproduced. \footnote{In Ref.~\cite{hadron99}, 
the values for parameters $a$ and $b$ have been presented 
by inappropriately multiplying $(2L+1)$ for each resonance 
with total angular momentum $L$. $a$ and $b$ lost their original
meanings. 
This problem has been corrected here. }

It should be pointed out, in this scheme
the roles played by the {\it t}-channel processes and 
the {\it s}- and {\it u}-channel have been highlighted.
We stress this feature since it is the prerequisite
for the motivation of studying resonance phenomena when other 
processes are present. 
Although the final determination of the two parameters $a$ and $b$ 
needs more constraints from experiment, 
it can still provide us 
with quantitative informations 
related to the nucleonic resonance excitations, especially in the polarization 
observables.

\subsection{ Cross sections}

Using the parameters determined through the above scheme, 
our first step is to study the cross section.

In Fig.~\ref{fig:(1)}, we re-studied the differential cross sections 
at four energy scales, $E_\gamma=1.225$, 1.45, 1.675 and 1.915 GeV. 
As a comparison, the exclusive contributions from the three channels 
are also presented. At forward angles, the pion exchange (dashed curves)
account for the steep forward peakings. 
The Pomeron exchange (dot-dashed curves) has quite small 
contributions at this energy region. However, we can see below, 
this small part will play important roles in some polarization observables. 

Notice that near threshold, the {\it s}- and 
{\it u}-channel contributions (dotted curves)
dominate over the {\it t}-channel 
$\pi^0$ exchange. This character can be justified 
by the precise measurements of the angular distributions.
Nevertheless, the polarization observables 
should be able to provide more rich informations relevant to the 
 the resonance interferences.

The multiplets 
of the representation 
$[{\bf 56, ^2 8}, 2, 2, J]$ in the quark model are found dominating 
over the other low-lying states in the cross sections. 
In the quark model, the resonances $P_{13}(1720)$ and $F_{15}(1680)$ 
are attributed to the $[{\bf 56, ^2 8}, 2, 2, J]$ representation
with $J=3/2$ and $5/2$ respectively 
in the harmonic oscillator basis, while the $P_{13}(1900)$
and $F_{15}(2000)$, to the $[{\bf 70, ^2 8}, 2, 2, J]$ representation.
With the same couplings of the constituent quark to the $\omega$
meson as other intermediate states, 
 the contributions 
from the resonances $P_{13}(1720)$ and $F_{15}(1680)$ are found much larger
than that from  
resonances $P_{13}(1900)$ and $F_{15}(2000)$. 
Comparing the two representations $[{\bf 56, ^2 8}, 2, 2, J]$ and 
$[{\bf 70, ^2 8}, 2, 2, J]$ to each other, we find 
two main features in their excitations lead to such a large difference.
The first feature is that the configuration of $[{\bf 56, ^2 8}, 2, 2, J]$
permittes its transitions to the nucleon groundstate
{\it via} the spin and orbital angular momentum splitting or 
non-splitting terms. 
Therefore, the three independent meson transition amplitudes 
$A^V_{1/2}$, $A^V_{3/2}$ and $S^V_{1/2}$ are all contributing.
However, for the transitions between the $[{\bf 70, ^2 8}, 2, 2, J]$ to the 
nucleon groundstate, the only non-vanishing amplitude 
is the $A^V_{1/2}$ due to the spin-splitting operator~\cite{prc98}. 
This feature suggests that states of $[{\bf 56, ^2 8}, 2, 2, J]$ have larger
transition space to $\omega N$ than those of $[{\bf 70, ^2 8}, 2, 2, J]$.
The second feature comes from the mass differences between
these two sets of resonances. 
The $P_{13}(1720)$ and $F_{15}(1680)$  have the masses 
close to the $\omega$ production threshold, thus,
they have larger absorption amplitudes than the $P_{13}(1900)$ and 
$F_{15}(2000)$ near threshold.
The dominant $P_{13}(1720)$ and $F_{15}(1680)$
can be seen significantly in the beam polarization observables.

In Fig.~\ref{fig:(2)}, the total cross sections 
for the three different processes are explicitly illustrated. 
Near threshold, 
the {\it s}- and {\it u}-channel contributions 
dominate over the {\it t}-channel $\pi^0$ exchange. 
As shown in the angular distributions, the 
cross section is dominated by the large angle mechanisms which 
are essentially governed by the resonance excitations.  
However, this contribution falls down quickly 
with the increasing energy. 
Above $E_\gamma=3$ GeV, the {\it t}-channel 
processes become dominant, especially at the forward angles. 
The energy-evolution of the cross sections indicates the change of the
relative strengths for these three mechanisms. 
As shown by the dashed and dot-dashed curves, above 3 GeV
the interferences between the Pomeron and $\pi^0$ exchanges will 
determine the forward angle behaviors for most observables. 
This feature provides us the opportunity
of fixing the parameters of the {\it t}-channels.

With the Pomeron exchange contribution, 
the model improves the calculations when the energy goes 
above $E_\gamma \sim 2.2$ GeV~\cite{prc98,FrimanSoyeur}.

\subsection{ Density matrix elements and parity asymmtry observables}

In vector meson photo-production, the final state vector meson decays 
($\omega\to \pi^+\pi^-\pi^0$)
provide access to the measurements of the vector meson decay 
density matrix elements, which can be related to the vector 
or tensor polarizations of the vector meson, or angular distributions 
of the cross sections~\cite{schilling}:
\begin{eqnarray}
\label{decay-distribution}
W(\cos \theta, \phi, \Phi) &=& W^0(\cos\theta, \phi, \rho^0_{\alpha\beta})
- P_\gamma \cos 2\Phi W^1(\cos\theta,\phi,\rho^1_{\alpha\beta}) \nonumber\\
& & -P_\gamma \sin 2\Phi W^2(\cos\theta,\phi,\rho^2_{\alpha\beta})\nonumber\\
& & + \lambda_\gamma P_\gamma W^3(\cos\theta,\phi,\rho^3_{\alpha\beta}) \ .
\end{eqnarray}
where
\begin{eqnarray}
W^0(\cos\theta, \phi, \rho^0_{\alpha\beta})&=&
\frac{3}{4\pi} [\frac 12 \sin^2\theta + \frac 12 (3\cos^2\theta -1)
\rho^0_{00}\nonumber\\
&&-\sqrt{2} \mbox{Re} \rho^0_{10}\sin 2\theta\cos\phi 
-\rho^0_{1-1}\sin^2\theta\cos 2\phi] \ , \\
W^1(\cos\theta,\phi,\rho^1_{\alpha\beta})&=& 
\frac{3}{4\pi} [\rho^1_{11} \sin^2\theta + \rho^1_{00}\cos^2\theta\nonumber\\
&&-\sqrt{2} \mbox{Re} \rho^1_{10}\sin 2\theta\cos\phi 
-\rho^1_{1-1}\sin^2\theta\cos 2\phi] \ , \\
W^2(\cos\theta,\phi,\rho^2_{\alpha\beta})&=&
\frac{3}{4\pi} [\sqrt{2} \mbox{Im} \rho^2_{10}\sin 2\theta\sin\phi
+ \mbox{Im}\rho^2_{1-1}\sin^2\theta\cos 2\phi ] \ , \\
W^3(\cos\theta,\phi,\rho^3_{\alpha\beta}) &=&
\frac{3}{4\pi} [ \sqrt{2} \mbox{Re}\rho^3_{10} \sin 2\theta\sin\phi
+ \mbox{Im} \rho^3_{1-1} \sin ^2\theta\sin 2\phi ] \ .
\end{eqnarray}
In the above equations, $\theta$ and $\phi$ are the polar and azimuthal 
angles of the normal direction of the $\omega$ decay plane with 
respect to the $z$-axis in the production plane, which in the helicity 
frame is defined as the direction of the $\omega$ meson momentum. 
$\Phi$ is the angle of the photon polarization vector with respect to 
the production plane.

In the helicity space, the density matrix elements can be explicitly 
expressed in terms of the 12 independent helicity amplitudes: 
\begin{eqnarray}  
\rho^0_{ik}&=  &\frac{1}{A}\sum_{\lambda\lambda_2\lambda_1}H_{\lambda_{v_i}  
\lambda_2,  
\lambda\lambda_1} H^*_{\lambda_{v_k}\lambda_2, \lambda\lambda_1},  
\nonumber\\
\rho^1_{ik}&=  &\frac{1}{A}\sum_{\lambda\lambda_2\lambda_1}  
H_{\lambda_{v_i}\lambda_2,  
-\lambda\lambda_1} H^*_{\lambda_{v_k}\lambda_2, \lambda\lambda_1},  
\nonumber\\  
\rho^2_{ik}&=  &\frac{i}{A}\sum_{\lambda\lambda_2\lambda_1}\lambda   
H_{\lambda_{v_i}\lambda_2, -\lambda\lambda_1} H^*_{\lambda_{v_k}\lambda_2,   
\lambda\lambda_1},\nonumber\\  
\rho^3_{ik}&=  &\frac{i}{A}\sum_{\lambda\lambda_2\lambda_1}\lambda   
H_{\lambda_{v_i}\lambda_2, \lambda\lambda_1} H^*_{\lambda_{v_k}\lambda_2,   
\lambda\lambda_1},   
\end{eqnarray}  
and   
\begin{equation}  
A=\sum_{\lambda_{v_i}\lambda\lambda_2\lambda_1} H_{\lambda_{v_i}\lambda_2,  
\lambda\lambda_1} H^*_{\lambda_{v_i}\lambda_2, \lambda\lambda_1},  
\end{equation}  
where $\rho_{ik}$ stands for $\rho_{\lambda_{v_i}\lambda_{v_k}}$,
 and $\lambda_{v_i}$, $\lambda_{v_k}$ denote the helicity of the 
produced vector mesons. $A$ is the cross section function.
With the linearly polarized photon beam, $\rho^0_{ik}$, 
$\rho^1_{ik}$ and $\rho^2_{ik}$ can be measured while with the circularly polarized 
photon beam, $\rho^0_{ik}$ and $\rho^3_{ik}$ can be measured. 
With the additional relations: 
$\rho^0_{00}+2 \rho^0_{11}\equiv 1$, which represents 
the normalized cross section, 
only 11 elements are linearly independent.

At high energies, the experimental data for vector meson 
photo-production exhibit characters of {\it t}-channel 
{\it natural}-parity exchange dominance and {\it s}-channel 
helicity conservation (SCHC). 
From high energy down to the resonance region, 
deviations from these features are expected due to the 
resonance excitations and the {\it t}-channel {\it unnatural}-parity
exchange. 
Such an interference due to the different relative strength and 
mechanisms 
can be illustrated by the density matrix elements. 
Concerning the {\it t}-channel {\it natural} and {\it unnatural}-parity 
exchanges, some model-independent features can be learned~\cite{schilling}: 
\begin{itemize}
\item Supposing only the {\it natural}-parity Pomeron exchange 
contributes in the diffractive phenomena, it can be justified that 
except that $\rho^0_{11}$, $\rho^1_{1-1}$ and $\mbox{Im} \rho^2_{1-1}$ 
have a value of 0.5, other density matrix elements vanish. 

\item For the exclusive {\it unnatural}-parity exchange,  
 one has $\rho^0_{11}=0.5$,
$\rho^1_{1-1}=-0.5$, and $\mbox{Im} \rho^2_{1-1}=-0.5$, while
the other elements will vanish.

\item In the $\pi^0$ exchange terms, the spin operator does not 
flip the helicity of the photon, therefore, the transition 
$\lambda_\gamma=\pm 1$ to $\lambda_v=0$ vanish. 

\end{itemize}

In the model for the Pomeron exchange, 
although the Pomeron behaves rather like a $C=+1$ isoscalar photon, 
those spin splitting terms will be significantly suppressed in the 
forward angles~\cite{phi99}, therefore, the transition 
$\lambda_\gamma=\pm 1$ to $\lambda_v=0$ ``vanish" in the 
forward angles, which makes the this process like a $0^{++}$ quantum number exchange
process. 
Consequently, 
the element $\rho^0_{00}$ vanishes in the forward 
direction due to the dominant {\it natural} or {\it unnatural}-parity
exchanges in the small momentum transfer.

In Fig.~\ref{fig:(3)}, ~\ref{fig:(4)}, and ~\ref{fig:(5)}, 
the density matrix elements $\rho^0_{ik}$, 
$\rho^1_{ik}$ and $\rho^2_{ik}$ involved in the 
measurement of Ref.~\cite{ballam73} 
are studied 
at small $t$ regions for $E_\gamma=2.8$, 4.7 and 9.3 GeV
from left to right in each row.
It should be pointed out, the non-relativistic resonance excitation 
model might meet trouble when applied to high energies, for example, 
$E_\gamma >$ 3 GeV. 
However, notice that this process decrease exponentially with the 
increasing energy, and becomes negligible around 3 GeV, 
 our study of the constraints to the {\it t}-channel
processes do not suffer troubles arising from this shortcoming, especially
at forward angles.

In Fig.~\ref{fig:(3)}, ~\ref{fig:(4)}, and ~\ref{fig:(5)},
the solid curves present the exclusive calculations for the {\it t}-channel 
reactions. The full calculations are illustrated
through the dashed curves. It shows that at $E_\gamma=2.8$ GeV, 
the resonance excitations have become negligible for most of the elements.
For $E_\gamma=9.3$ GeV, the dashed curves can not be distinguished 
from the solid ones.
Here, although the data~\cite{ballam73} for the density matrix elements 
are very sparse, they provide an interesting test of the model 
of {\it t}-channel $\pi^0$ and Pomeron exchanges, which will essentially
determine the forward angle behaviors of the density matrix elements.

As summarized in former paragraph, 
the non-zero elements $\rho^0_{11}$, $\rho^1_{1-1}$
and $\mbox{Im}\rho^2_{1-1}$ will reflect the interferences 
between these two processes. 
In Fig.~\ref{fig:(3)}, the calculations of the elements $\rho^0_{00}$, $\rho^0_{1-1}$
and $\mbox{Re}\rho^0_{10}$ are in good agreement with the 
data for small $t$. 
There is no datum available for the
$\rho^0_{11}$ in experiment. An interesting feature for this element 
is that it 
will have the value 0.5 at forward  angles 
for both {\it natural} and {\it unnatural}-parity exchanges. 
The close-to-0.5 value we find might suggest that 
this element does not involve strong 
interferences within these two {\it t}-channel processes.

Comparing the $\rho^1_{1-1}$ at these three energies in Fig.~\ref{fig:(4)}, 
one can see that 
the data have been reproduced impressively.  
The relative strengths between the $\pi^0$ and Pomeron exchanges
as well as the energy-evolutions for these two terms 
are essential for reproducing this element. 
At 2.8 GeV, the {\it unnatural}-parity exchange dominates over the 
{\it natural}-parity one, thus 
produces negative values for this element. 
With the increasing energy,  
the {\it natural}-parity exchange becomes dominant. 
Therefore, at 4.7 GeV, 
positive values for the $\rho^1_{1-1}$ are found. 
At 9.3 GeV, 
the value reaches 0.5, which suggests that the Pomeron exchange has
absolutely dominated over other processes. 

For most of the elements, our calculations are in good agreement 
with the measurement except for the $\mbox{Im}\rho^2_{1-1}$ at 9.3 GeV
in Fig.~\ref{fig:(5)}, 
where large discrepancies are found. 
Interestingly, as summarized at the beginning of this section, 
at 9.3 GeV, the {\it natural}-parity exchange becomes dominant,  
therefore, it will have the value 0.5 at forward angles. 
The calculation consistently output this feature, while however, 
the measurement provides small negative values. 
The data here can not be explained by the influences from the 
{\it unnatural}-parity exchange and the {\it s}- and 
{\it u}-channel resonance excitations due to their small
cross section at this energy region.  
A precise measurement for this element should be important 
to clarify the discrepancies.

Concerning the mechanism of the {\it t}-channel parity exchanges, 
one can investigate the parity asymmetry observable $P_\sigma$, 
which can be related to two of the density matrix elements,
 $\rho^1_{1-1}$ and $\rho^1_{00}$:
\begin{equation}
P_\sigma=2 \rho^1_{1-1}-\rho^1_{00} \ .
\end{equation}
These two elements can be determined 
in the linearly polarized photon beam measurement. 
As discussed above, 
in the pure {\it natural}-parity exchange process, 
$\rho^1_{1-1}=0.5$ and $\rho^1_{00}=0$ will result in 
$P_\sigma=+1$. Similarly, if the {\it unnatural}-parity exchange 
process contributes exclusively, one will get $P_\sigma=-1$
with $\rho^1_{1-1}=-0.5$ and $\rho^1_{00}=0$. 
At high energies and at the forward angles, 
the deviation of the parity asymmetries from  $+1$ or $-1$ will 
suggest that both {\it natural} and {\it unnatural}-parity 
exchanges exist in the reaction. 
However, such a conclusion can not be applied to the low energy region 
due to the presence of resonance excitations, which will violate
the {\it s}-channel helicity conservation, and shift 
the parity asymmetry away from $+1$ or $-1$. 
In other words, if one expects to disentangle informations relevant to 
the resonance excitations, the prerequisite is to consistently 
include the {\it natural}-parity exchange interference.

The results for $P_\sigma$ at $E_\gamma=2.8$, 4.7 and 9.3 GeV 
are shown in Fig.~\ref{fig:(6)}. 
It shows that at 2.8 GeV, the {\it unnatural}-parity $\pi^0$ 
exchange is dominant, thus negative asymmetries are produced. 
With the increasing energy, the {\it natural}-parity 
Pomeron exchange becomes more and more important and 
then the asymmetry will change to positive. 
At 9.3 GeV, the $\pi^0$ and 
resonance excitations become negligible and the dominant
Pomeron exchange produces the parity asymmetries very close to $+1$.
Apparently, the parity asymmetry at high energies and forwards angles
shows great sensitivities to 
the underway {\it natural} or {\it unnatural}-parity exchange
mechanisms.

The dashed curves illustrate the effects if the {\it s}- and {\it u}-channel 
resonance excitations are taken into account. One can see that 
their effects at small $t$ is quite negligible. 
It confirms the scheme to determine the {\it t}-channel parameters
at higher energies, and
then apply them to the resonance region.

\subsection{ Beam polarization observables}
 
The polarization observable asymmetries becomes interesting since 
contributions from individual resonances might be amplified 
in the polarization observables, therefore, produce a detectable
effects. 
In comparison with the density matrix elements, 
here one select the experimental situation by polarizing 
the spin of the initial and/or final state particles.

In particular,
with the linearly polarized photon beam, the first three terms 
in Eq.~\ref{decay-distribution} can be measured. In the final state
$\omega$ meson decay plane, one can select $\Phi=0^\circ$, thus,
the third term in Eq.~\ref{decay-distribution} vanishes and only 
$W^0$ and $W^1$ have contributions. At the direction $\phi=90^\circ$
and $\theta=90^\circ$, one obtains: 
\begin{eqnarray}
W(\cos 90^\circ, 90^\circ, 0^\circ )&=&\frac{3}{4\pi} 
\left [\frac 12 - 
\frac 12 \rho^0_{00} + \rho^0_{1-1} \right ]\nonumber\\
&&-P_\gamma \frac{3}{4\pi}
\left [ \rho^1_{11} + \rho^1_{1-1} \right ] \ .
\end{eqnarray}

With the relation $2\rho^0_{11}+\rho^0_{00}=1$, 
the beam polarization asymmetry is defined as:
\begin{equation}
\label{beam-asymm-A}
\check{\Sigma}_A\equiv\frac{\rho^1_{11}+\rho^1_{1-1}}
{\rho^0_{11}+\rho^0_{1-1}} = \frac{\sigma_\parallel -\sigma_\perp}
{\sigma_\parallel +\sigma_\perp} \ ,
\end{equation}
where $\sigma_\parallel$ represents the pion cross section of the $\omega$
meson decay with the pions submerged in the photon polarization plane, 
while $\sigma_\perp$ represents that with the pions perpendicular to it.

In Fig.~\ref{fig:(7)}, $\check{\Sigma}_A$ is investigated 
for four energy scales, i.e. $E_\gamma=1.15$, 1.25, 1.35 and 1.45 GeV 
within the energy access of the GRAAL Collaboration.
The solid curves denote the full calculations, while the 
dashed curves denote the exclusive calculations with the {\it t}-channel 
processes. The dotted curves are obtained by eliminating the 
Pomeron exchange in the full calculations.

Note that, for the pure {\it natural}-parity exchange process, 
one has $\check{\Sigma}_A=+1$, while for the pure {\it unnatural}-parity
exchange, $\check{\Sigma}_A=-1$.
~\footnote{From $\check{\Sigma}_A=\pm 1$, one can not directly reach 
such a conclusion that the exclusive 
{\it natural} or {\it unnatural}-parity
exchange have happened in the reaction, since the asymmetries of the 
double-flip amplitudes can also lead to the same result. }

Here, several lessons can be learned:
\begin{itemize}
\item 
As shown by the dashed curves,  
the negative asymmetries are due to the dominant $\pi^0$ exchange
at low energy regions. 

\item Comparing the dashed curves to the solid ones, 
one can see that the influences from the {\it s}- and {\it u}-channels 
are very significant, especially near threshold. 

\item The effects from the {\it natural}-parity exchange
are shown explicitly at forward angles by the dotted curves. 
This process has very small cross sections at low energies, therefore, 
can not be seen clearly in the angular distributions. 
However, it intervenes strongly at forward angles in $\check{\Sigma}_A$
and deviates the asymmetries from $-1$. This feature confirms our 
argument that a reliable description of the {\it t}-channel 
processes should be important in the study of the polarization observables
concerning the resonance excitations.

\item The large angle asymmetries are dominated by the {\it s}- and 
{\it u}-channel processes, in which signals for  
some individual resonances can be derived.

\end{itemize}

In the $\omega$ meson production, the most interesting resonance 
should be the $P_{13}(1720)$, of 
which the coupling to the $\omega N$ channel has not been established
 in experiment. 
As summarized in the previous subsection, the relatively 
large contributions from the $P_{13}(1720)$ are due to 
its NRCQM configuration and near-threshold mass. 

However,   
according to the PDG~\cite{PDG98}, the $P_{13}(1720)$ 
 couples stronger to the isovector 
$\rho$ rather than the isoscalar $\omega$ meson. 
A question arising here should be, ``if one can find signals for the 
presence of the $P_{13}(1720)$ given the beam polarization data available?" 

In this calculation, we find that the beam polarization asymmetries 
are also very sensitive to the $P_{13}(1720)$ excitation. 
Supposing the $P_{13}(1720)$ has no coupling to the $\omega N$ channel, 
we present the $P_{13}(1720)$-absent
asymmetries with the dot-dashed curves in Fig.~\ref{fig:(7)}. 
Very significant effects are found at backward angles, while
the influences at forward angles are found negligible. 
Such an effect can be tested by the measurements of GRAAL and CLAS.

With $\phi=0^\circ$, $\theta=90^\circ$ and $\Phi=0^\circ$ 
in Eq.~\ref{decay-distribution}, 
one can define another observable:
\begin{equation}
\label{beam-asymm-B}
\check{\Sigma}_B=\frac{\rho^1_{11}-\rho^1_{1-1}}{\rho^0_{11}-\rho^0_{1-1}} \ .
\end{equation}
It can be easily seen that for the exclusive {\it natural} or 
{\it unnatural}-parity exchanges, $\check{\Sigma}_B$ has values 
$-1$ or $+1$, respectively, which are opposite to the cases 
for $\check{\Sigma}_A$.  
When both processes exist in the reaction, a deviation from $+1$ or $-1$
is expected. 

In Fig.~\ref{fig:(8)}, the full calculations 
of $\check{\Sigma}_B$ (solid curves) are presented,
in which the angular distributions show drastic changes. 
In comparison with Eq.~\ref{beam-asymm-A}, 
due to the possible cancellations in the denominator of $\check{\Sigma}_B$, 
this observable 
might be sensitive to such effects that are not obvious 
in $\check{\Sigma}_A$.
In fact, we find that the ``defect" of the Pomeron exchange model 
can be see clearly in $\check{\Sigma}_B$ when the exclusive Pomeron exchange
is calculated. For the Pomeron exchange model, $\check{\Sigma}_A=+1$ 
is obtained with a negligible deviation, while for $\check{\Sigma}_B$,
large deviations from $-1$ at middle angles are found.
Although the dominant resonance amplitudes can merge such a 
``defect" efficiently, this observable should be more model-dependent 
than $\check{\Sigma}_A$.

In Ref.~\cite{tabakin}, the beam polarization is defined as:
\begin{equation}
\check{\Sigma}\equiv  \overrightarrow{\Sigma}\cdot{\hat x}{\cal T}
=\frac{2\rho^1_{11}+\rho^1_{00}}{2\rho^0_{11}+\rho^0_{00}} \ ,
\end{equation}
where $2\rho^0_{11}+\rho^0_{00}=1$ represents 
the normalized cross section function. 
In this case, the derivation of $\check{\Sigma}$ in experiment
requires the integration over angle $\phi$. 
In our previous investigations~\cite{plb98,prc98,hadron99},
the beam polarization $\check{\Sigma}$ was studied. 

Some common features can be derived for the {\it natural} 
and {\it unnatural}-parity exchanges:

\begin{itemize}
\item For the exclusive {\it natural} or {\it unnatural}-parity exchanges, 
$\check{\Sigma}$ vanishes due to $\rho^1_{11}$ and $\rho^1_{00}=0$. 
~\footnote{Note that, 
the Pomeron, which is rather like an isoscalar photon, 
will produce a negligible, but non-zero asymmetry in $\check{\Sigma}$.} 

\item With the $\pi^0$ plus Pomeron exchanges, this observable
almost vanishes since no interferences exist between the real-amplitude 
$\pi^0$ and imaginary-amplitude Pomeron exchanges. 

\item As a direct deduction of the previous two points, 
large asymmetries can only be produced by other non-diffractive 
contributions. 

\end{itemize}

In Fig.~\ref{fig:(9)}, the calculations of $\check{\Sigma}$ are presented. 
The solid curves denote the full calculations, while
the dashed curves are for the calculations without the {\it s}- 
and {\it u}-channel resonance contributions. 
We also present the Pomeron-absent results through the dotted curves, 
of which however, the changes are not as drastic as in $\check{\Sigma}_A$.
The $P_{13}(1720)$-absent effects are also shown by the dot-dashed curves.

In comparison with the preliminary study~\cite{hadron99}, 
more quantitative results are obtained here through
the extensive investigation over the available data.

\section{ Conclusions and discussions} 

In this work, we studied three contributing processes in the $\omega$ meson 
photo-production near threshold. In comparison with the previous 
approach, here, the introduction of the {\it t}-channel {\it natural}-parity
Pomeron exchange improves the model predictive power. 
Although the Pomeron exchange has only small amplitudes 
near threshold, which does not produce significant effects in the 
differential cross sections, it has strong interferences with the 
$\pi^0$ exchange terms at forward angles in the beam polarization asymmetries.
Taking the advantage of high energy measurement, 
the Pomeron exchange can be determined. 
Consequently, the interferences between the Pomeron and $\pi^0$ 
exchanges can be studied with respect to the intermediate-high-energy 
measurement at small $t$~\cite{ballam73}, through which the 
$\pi^0$ exchange can be determined. 
The reliable constraints found for the {\it t}-channels 
highlight the roles played by the resonance excitations. 
We stress that such a constraint is the prerequisite for the study 
of resonance excitations in various observables.

Proceeding to the study of the {\it s}- and {\it u}-channel resonance 
excitations in the beam polarization observables, we find that 
 the multiplets of 
representation $[{\bf 56, ^2 8}, 2, 2, J]$, i.e. $P_{13}(1720)$ and $F_{15}(1680)$ 
have large contributions in this reaction. 
The role played by the 
$P_{13}(1720)$ is analyzed in the beam polarization asymmetries. 
This resonance
becomes interesting due to its mass position near threshold 
and relatively larger phase space in the NRCQM. 
Since its coupling to the $\omega N$ has not been known in experiment,
the forthcoming data from GRAAL for the beam polarization asymmetries
will be able to establish the nature of this resonance 
in the $\omega N$ channel.

\acknowledgments
The author thanks J.-P. Didelez, M. Guidal and
E. Hourany for useful discussions concerning the GRAAL experiments. 
Fruitful discussions with Z.-P. Li and B. Saghai are acknowledged.
Useful communications with T.-S.H. Lee are acknowledged.

This work was supported in part by the 
``Bourses de Recherche CNRS-K.C. WONG" and IPN-Orsay.

%
%
%
%
%
%
\begin{figure}
\caption{ Differential cross sections for
$\gamma p \to \omega p$. The solid curves denote the full calculations, 
while the dashed, dotted, and dot-dashed curves denote the 
exclusive calculations of the $\pi^0$ exchange, {\it s}- and {\it u}-channels, 
and Pomeron exchange, respectively.
The data are from the SAPHIR Collaboration~\protect\cite{klein}. }
\protect\label{fig:(1)}
\end{figure}

\begin{figure}
\caption{ Total cross sections for $\gamma p\to \omega p$. 
Notations are the same as Fig.~\ref{fig:(1)}. 
Data comes from ~\protect\cite{klein} (full dot) 
and ~\protect\cite{ABBHHM68,ballam73,old-data} 
(empty square). 
 }
\protect\label{fig:(2)}
\end{figure}

\begin{figure}
\caption{ Density matrix elements $\rho^0_{ik}$. 
From left to right for each row, the photon energy $E_\gamma$ 
will be 2.8, 4.7, and 9.3 GeV, respectively. 
The solid curves denote the calculations of 
the {\it t}-channel $\pi^0$ and Pomeron exchange. 
The dashed curves denote the full calculations. 
Data are from ~\protect\cite{ballam73}.
 }
\protect\label{fig:(3)}
\end{figure}

\begin{figure}
\caption{ Density matrix elements $\rho^1_{ik}$. 
The notations are the same as Fig.~\ref{fig:(3)}.
 }
\protect\label{fig:(4)}
\end{figure}
\begin{figure}
\caption{ Density matrix elements $\rho^2_{ik}$. 
The notations are the same as Fig.~\ref{fig:(3)}.
 }
\protect\label{fig:(5)}
\end{figure}
\begin{figure}
\caption{ Parity asymmetry $P_\sigma$. 
The notations are the same as Fig.~\ref{fig:(3)}.
 }
\protect\label{fig:(6)}
\end{figure}
\begin{figure}
\caption{ Beam polarization asymmetry $\check{\Sigma}_A$ 
at four energy scales. 
The solid curves denote the full calculations, while the dashed curves 
denote results of the {\it t}-channel
$\pi^0$ and Pomeron exchange, and the dotted curves, of 
$\pi^0$ plus {\it s}- and {\it u}-channels. The dot-dashed curves 
denotes the $P_{13}(1720)$-absent effects in the full calculations.
 }
\protect\label{fig:(7)}
\end{figure}
\begin{figure}
\caption{ Beam polarization observable $\check{\Sigma}_B$ 
at four energy scales. 
The solid curves denote the full calculations, while the dot-dashed 
curves denote the $P_{13}(1720)$-absent effects in the full calculations.
 }
\protect\label{fig:(8)}
\end{figure}
\begin{figure}
\caption{ Beam polarization observable $\check{\Sigma}$ 
at four energy scales. The notations are the same as Fig.~\ref{fig:(7)}.
 }
\protect\label{fig:(9)}
\end{figure}

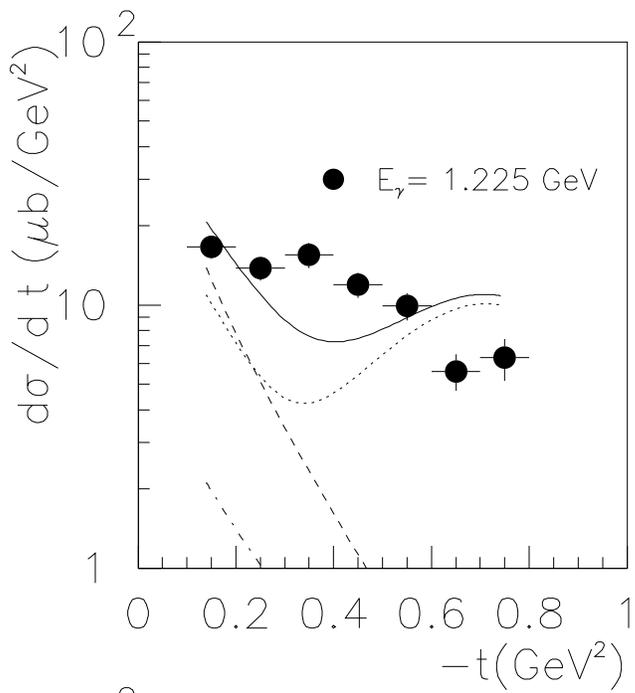
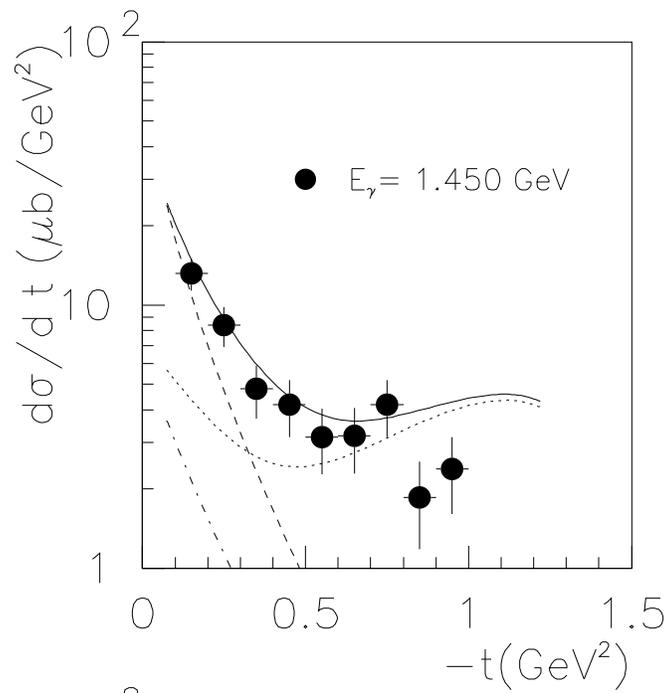
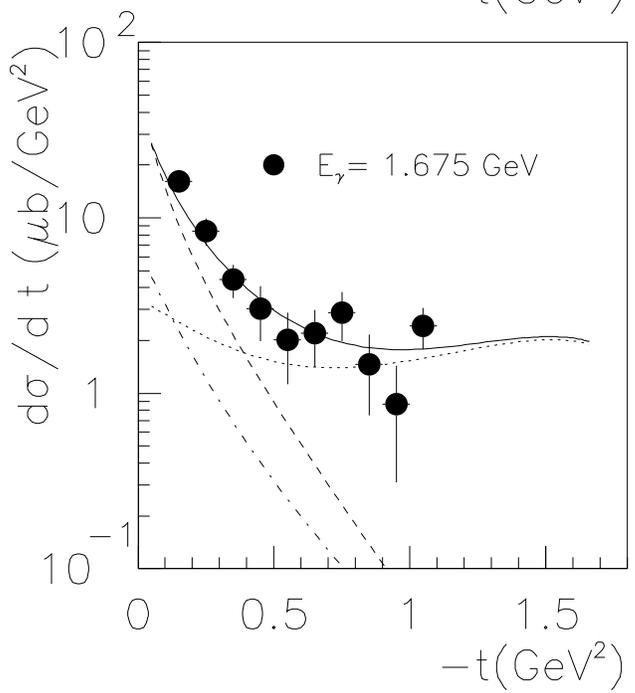
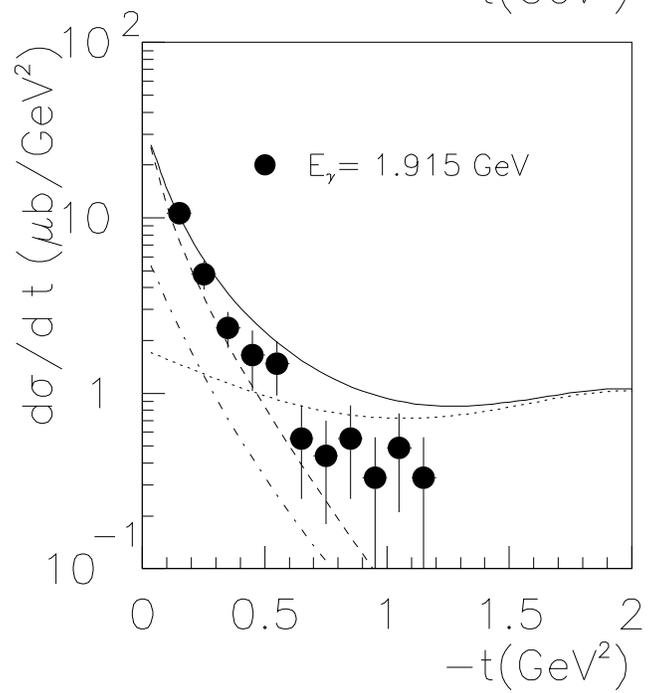
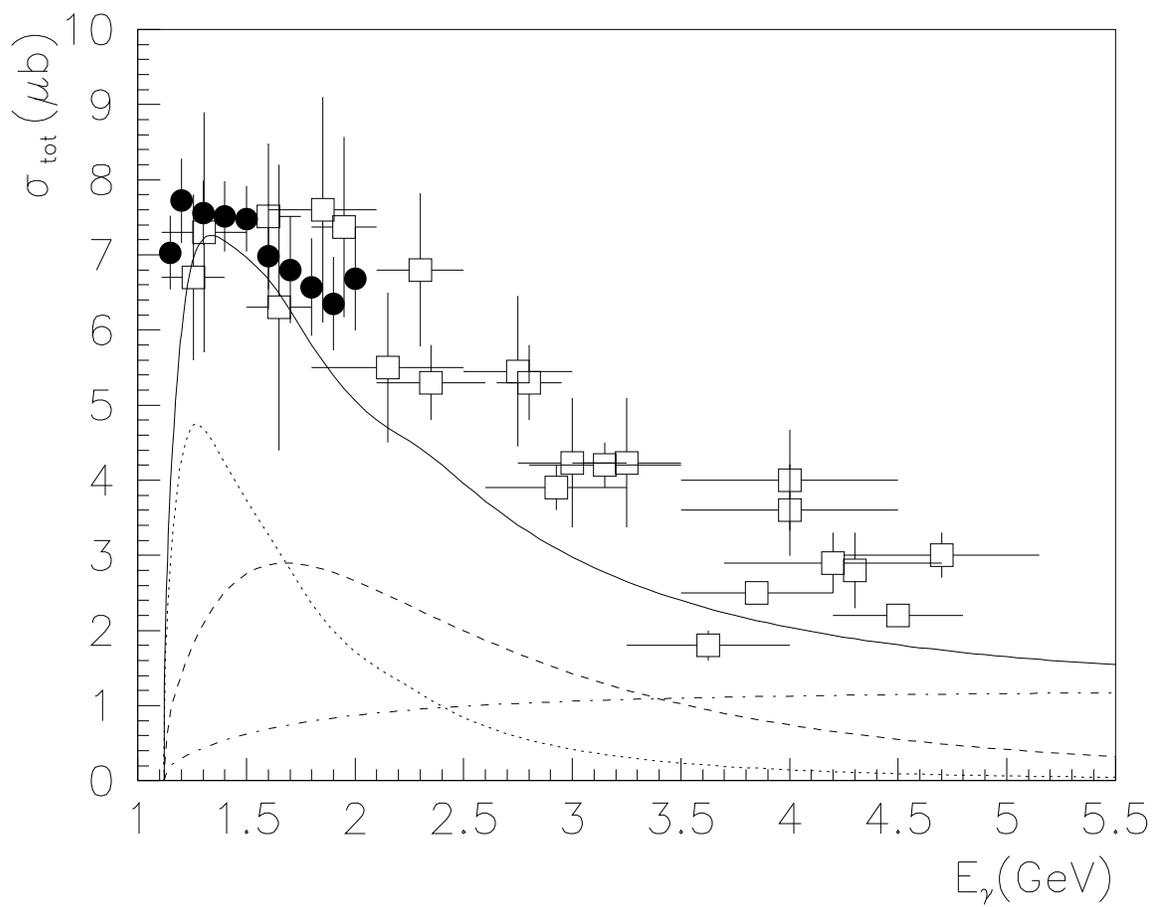
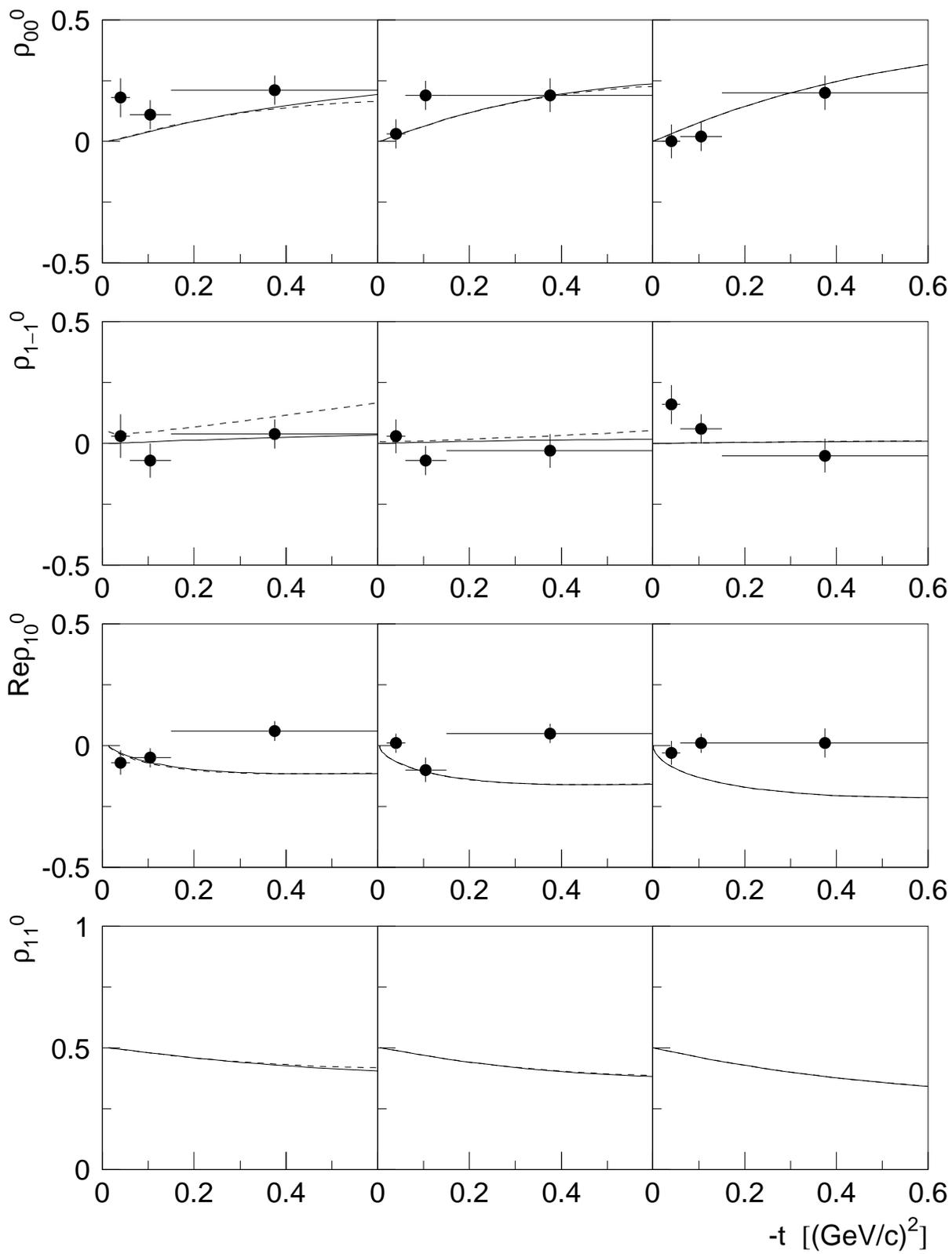
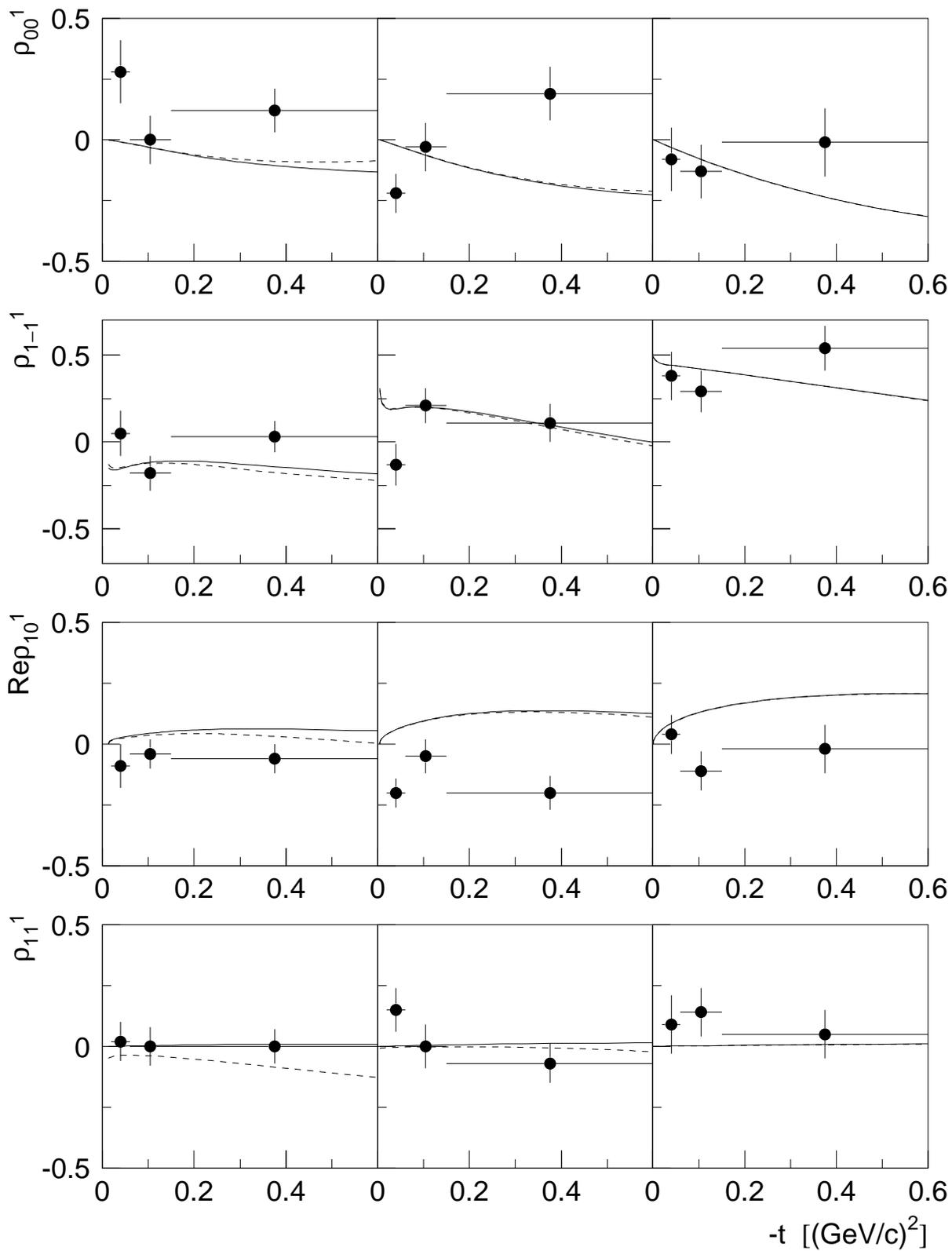
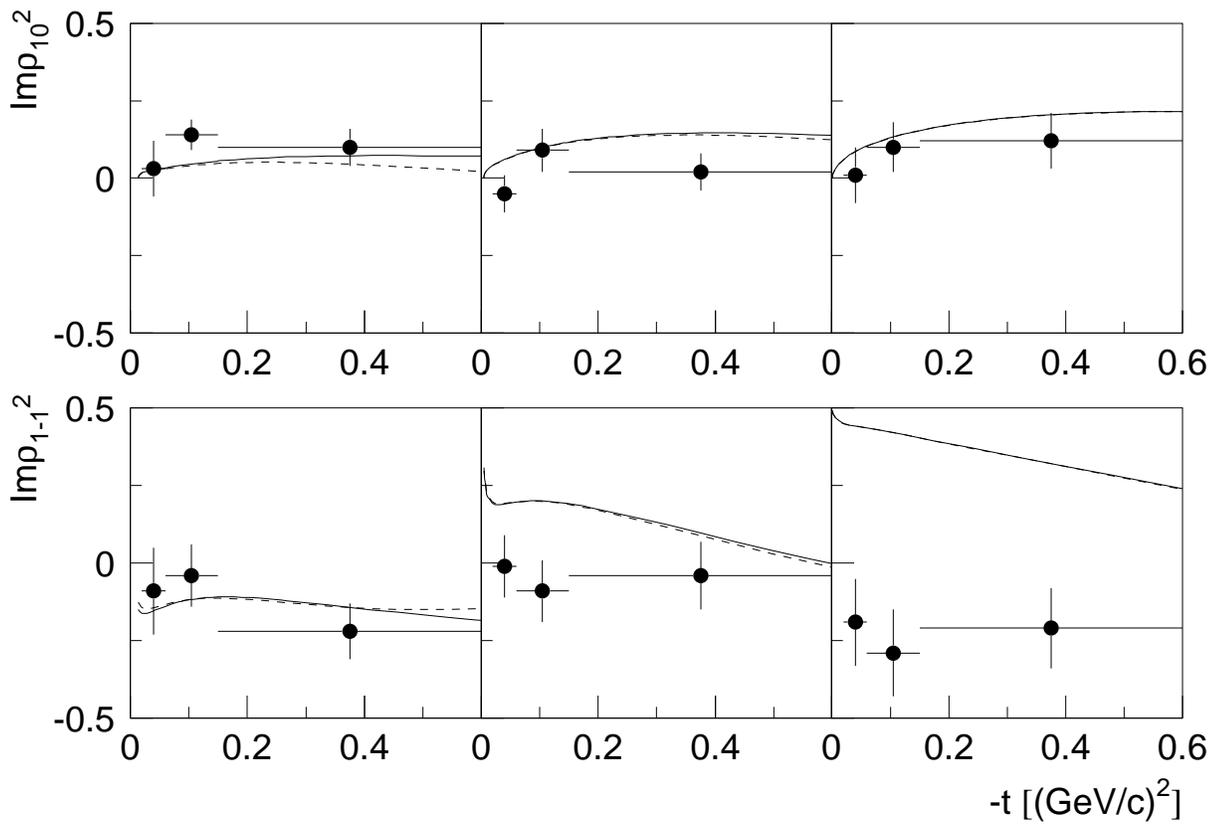
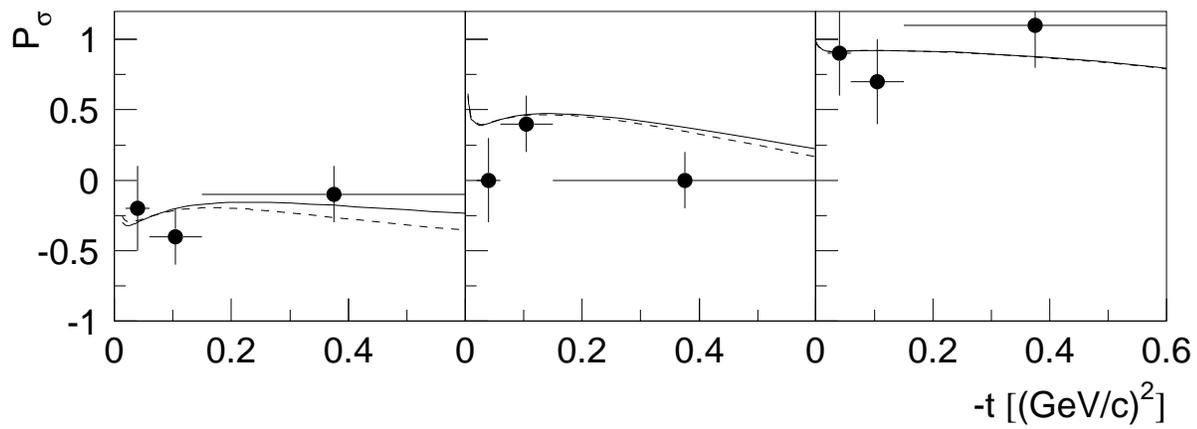
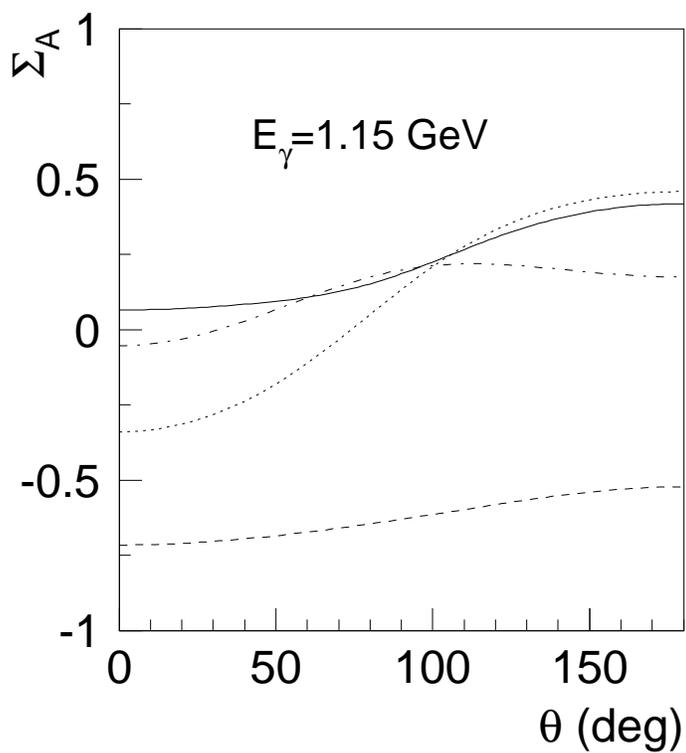
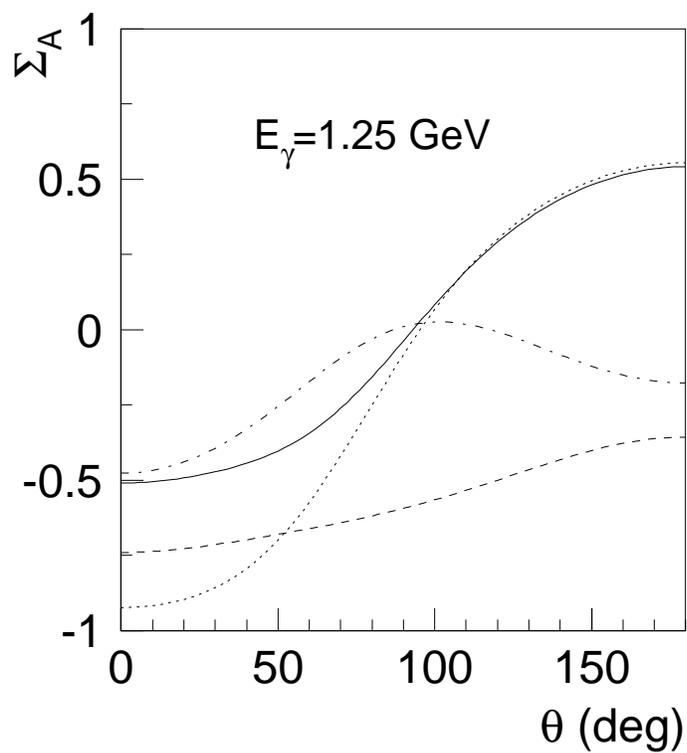
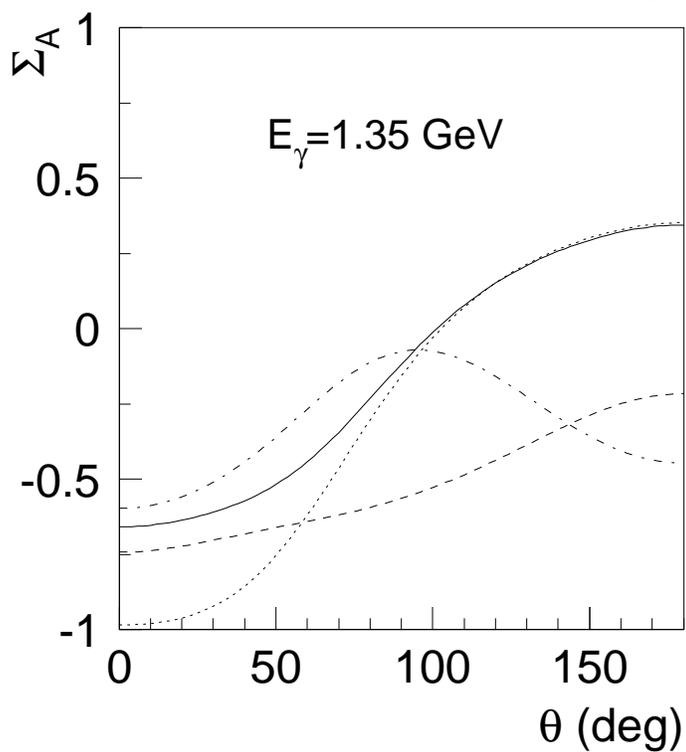
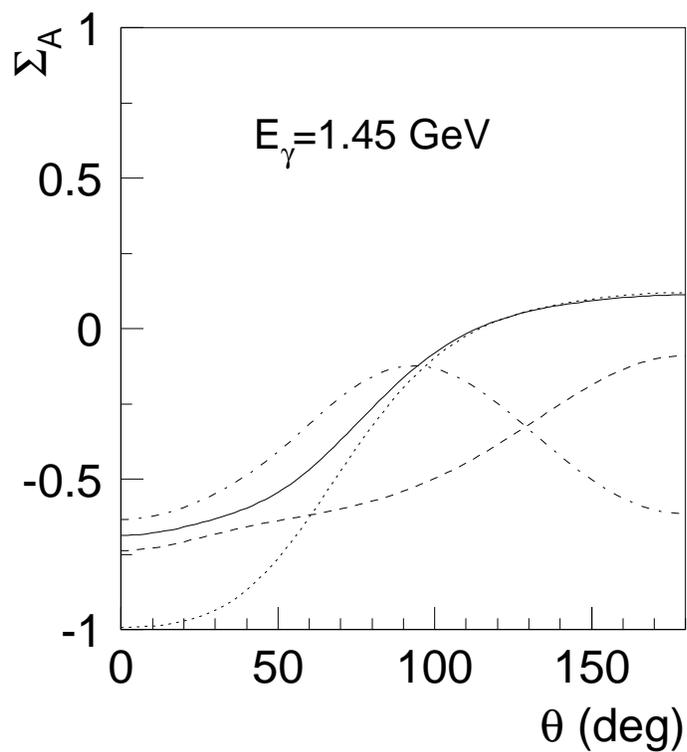
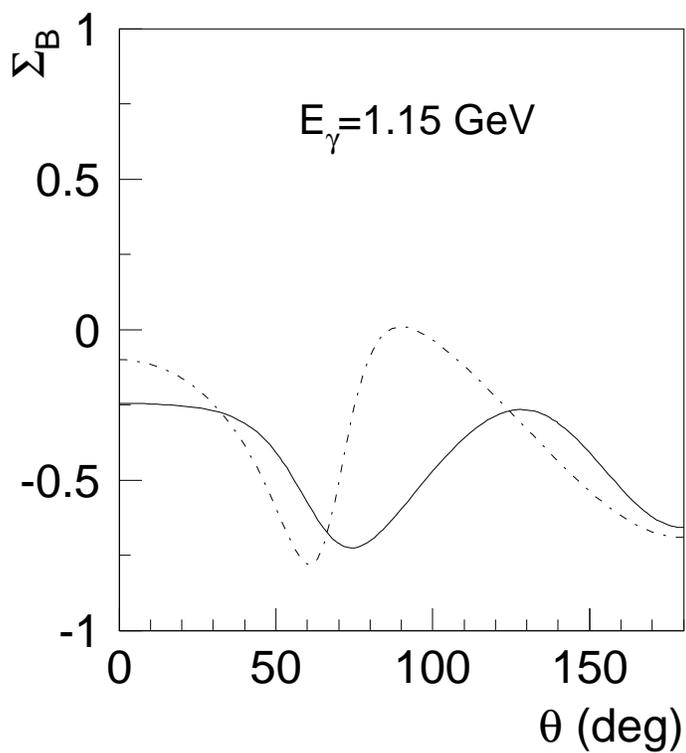
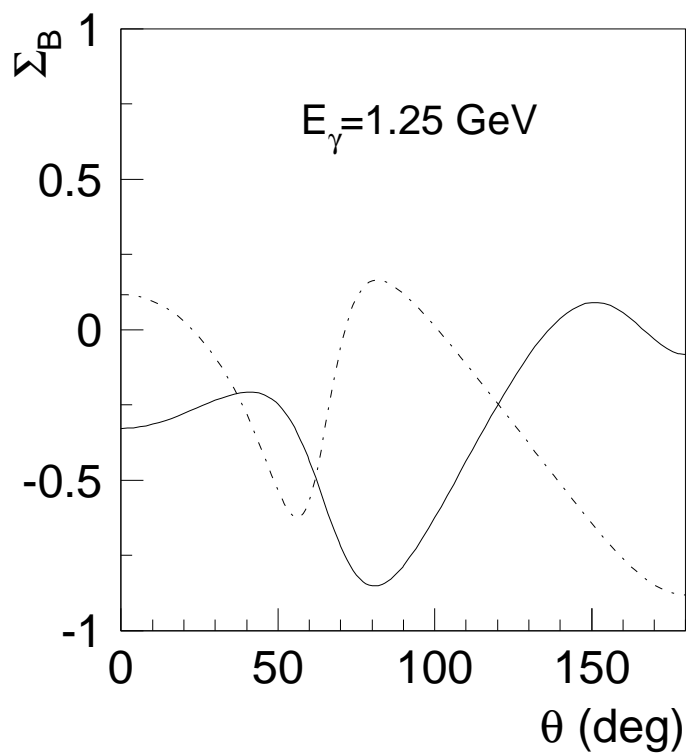
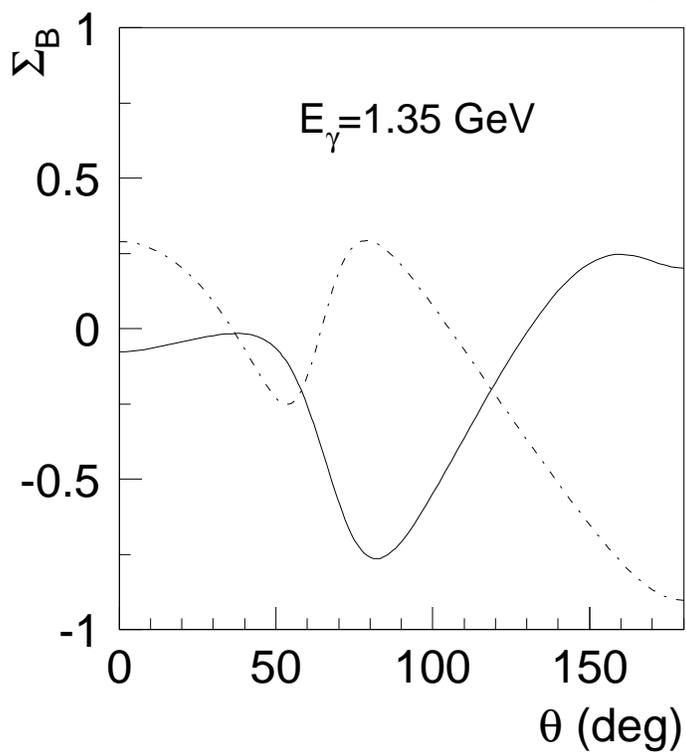
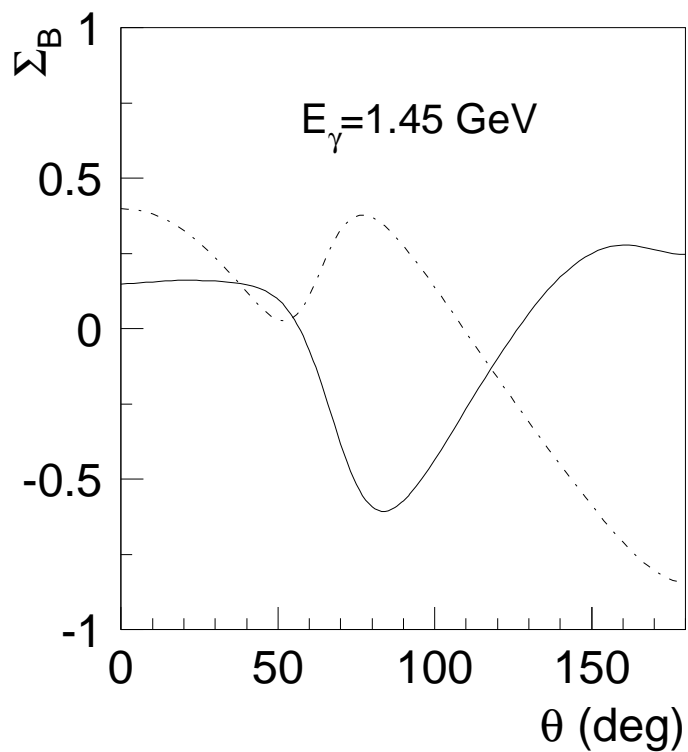
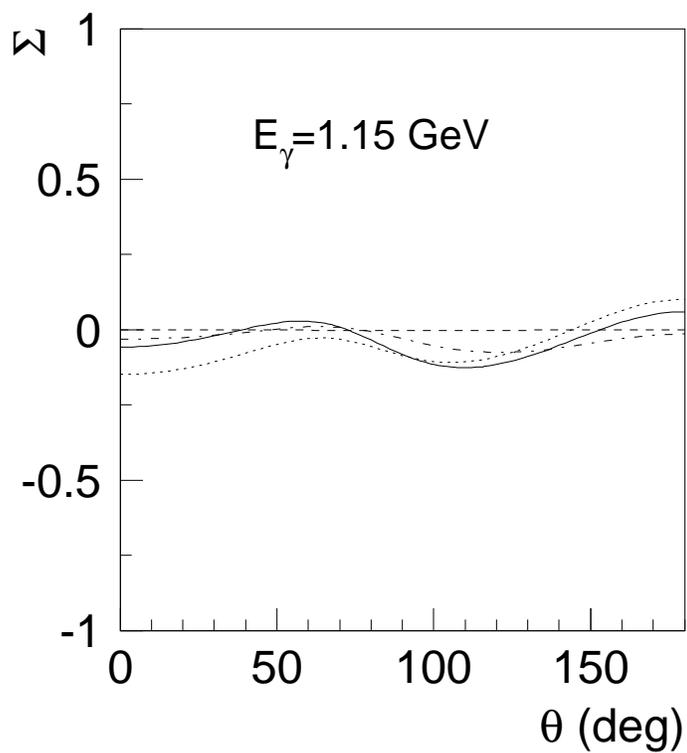
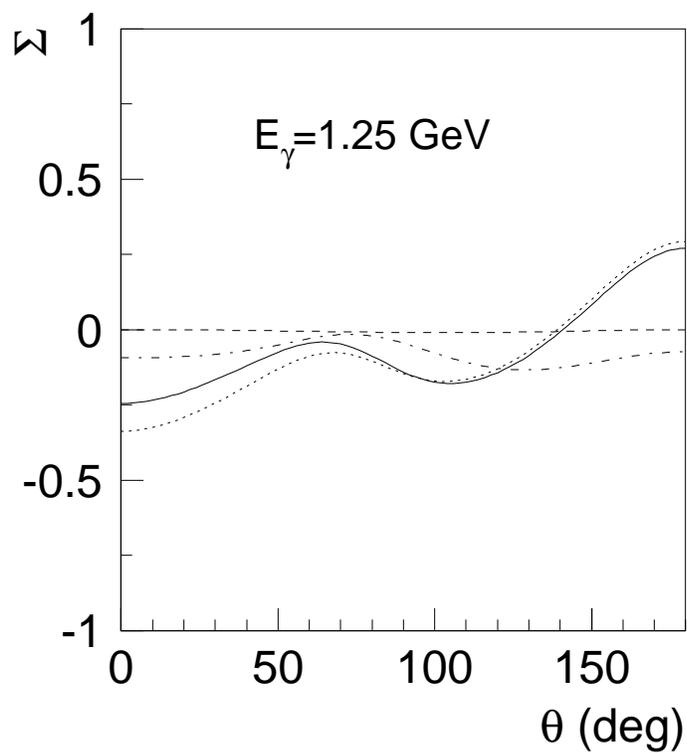
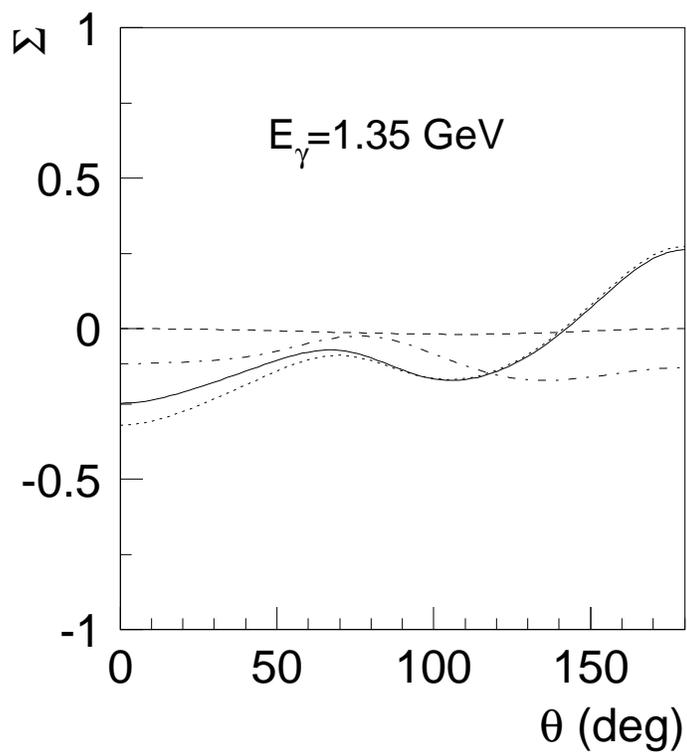
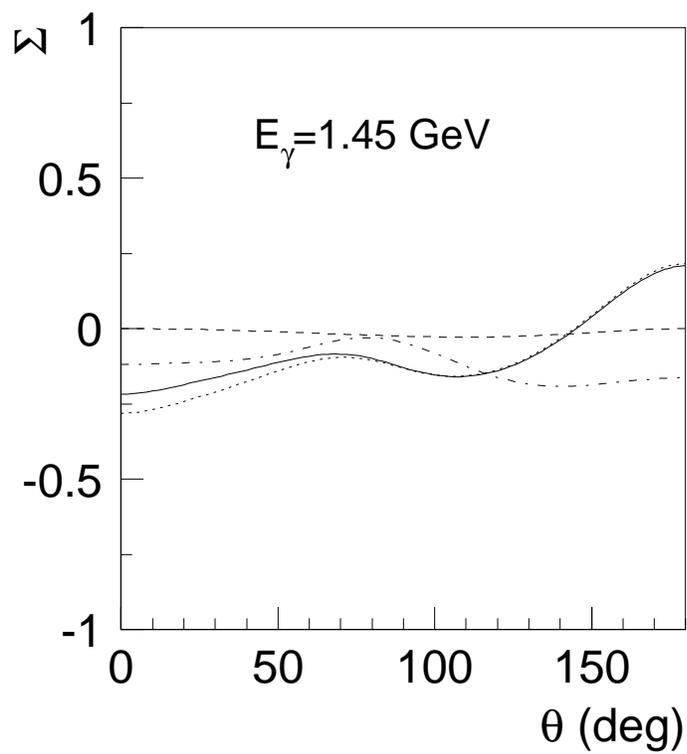
\end{document}